\documentclass{article}
\pdfoutput=1
 
\newif\ifIEEE
\IEEEfalse

\usepackage{arxiv}

\usepackage[utf8]{inputenc} 
\usepackage[T1]{fontenc}    
\usepackage{hyperref}       
\usepackage{url}            
\usepackage[nolist]{acronym}
\usepackage{graphicx}
\usepackage{subfig}
\usepackage{amsmath}
\usepackage{url}
\usepackage{booktabs}
\usepackage{multirow}
\usepackage{mathtools}

\makeatletter
\newcommand\footnoteref[1]{\protected@xdef\@thefnmark{\ref{#1}}\@footnotemark}
\makeatother

\begin{acronym}
    \acro{VR}[VR]{Virtual Reality}
\acro{CNN}[CNN]{Convolutional Neural Network}
\acro{CRNN}[CRNN]{Convolutional Recurrent Neural Network}
\acro{FF}[FF]{Feed Forward}
\acro{DCASE}[DCASE]{Detection and Classification of Acoustic Scenes and Events}
\acro{GRU}[GRU]{Gated Recurrent Unit}
\acro{ReLU}[ReLU]{Rectified Linear Unit}
\acro{tSNE}[t-SNE]{t-Distributed Stochastic Neighbour Embedding}
\acro{GA}[GA]{Genetic Algorithm}
\acro{AIR}{Acoustic Impulse Response}
\acro{MINC}[MINC]{Materials in Context Database}
\acro{DNN}{Deep Neural Network}
\acro{FIR}{Finite Impulse Response}
\acro{IIR}{Infinite Impulse Response}
\acro{VRC}[VRC]{Variance Ratio Criterion}
\acro{RNN}{Recurrent Neural Network}
\acro{SED}[SED]{Sound Event Detection}
\acro{ISO}{Intl. Organization for Standardization}
\acro{TOA}{Time of Arrival}
\acro{SVM}{Support Vector Machine}
\acro{ARMA}{Autoregressive Moving Average}
\acrodefplural{TOA}[ToAs]{Times-of-Arrival}

\end{acronym}

\title{Detecting Sound-Absorbing Materials in a Room from a Single Impulse Response using a CRNN}

\author{
  Constantinos Papayiannis\\
  Dept. of Electrical and Electronic
Engineering\\
Imperial College London\\
London SW7 2AZ, U.K. \\
  \texttt{papayiannis@imperial.ac.uk} \\
   \And
 Christine Evers\\
  Dept. of Electrical and Electronic
Engineering\\
Imperial College London\\
London SW7 2AZ, U.K. \\
  \texttt{c.evers@imperial.ac.uk} \\
     \And
  Patrick A. Naylor\\
  Dept. of Electrical and Electronic
Engineering\\
Imperial College London\\
London SW7 2AZ, U.K. \\
  \texttt{p.naylor@imperial.ac.uk}
}

\begin{document}
\maketitle

\begin{abstract}
The materials of surfaces in a room play an important room in shaping the auditory experience within them. Different materials absorb energy at different levels. The level of absorption also varies across frequencies. This paper investigates how cues from a measured impulse response in the room can be exploited by machines to detect the materials present. With this motivation, this paper proposes a method for estimating the probability of presence of 10 material categories, based on their frequency-dependent absorption characteristics. The method is based on a CNN-RNN, trained as a multi-task classifier. The network is trained using a priori knowledge about the absorption characteristics of materials from the literature. In the experiments shown, the network is tested on over 5,00 impulse responses and 167 materials. The F1 score of the detections was 98\%, with an even precision and recall. The method finds direct applications in architectural acoustics and in creating more parsimonious models for acoustic reflections.

\end{abstract}


\section{Introduction}

In \cite{Dokmanic2011}, the question was asked ``Can one hear the shape of a room?''. The question led to  a discussion on the task of geometry estimation and  subsequent interest in the field \cite{Moore2013}. It was shown that the \ac{AIR} encodes information   that enables the inference of the shape of the room. In this paper, we ask a similar question, which is: Can one hear the materials in a room?. The materials of surfaces in a room play an essential  part in shaping the auditory experience within them. As sound  interacts with surfaces, energy is absorbed and the level of absorption depends on the material of the surface  \cite{Kuttruff2009}. The level of absorption does not only depend on the material but also on the frequency of the sound. The aim of this work is to be able to detect the presence of materials in an acoustic environment from a single \ac{AIR}, based on these frequency-dependent absorptions.  This is in contrast to other methods that require the extraction of samples of  materials from the environment, such as  \acs{ISO}-354 and \acs{ISO}-10534. Detecting the present materials  finds direct applications in modeling for architectural acoustics and also paves the way for more parsimonious material-aware representations of reflections in \acp{AIR} \cite{Papayiannis2018Thesis}.

Related work was motivated by the need to recreate the acoustics of a room. Knowledge of how sound is absorbed in an enclosure improves computer models of it. The models enable a study of the room's acoustics through  simulations and help to find ways to improve its acoustics by altering its architecture. In \cite{Arteaga2013}, a method  based on \acp{GA} was proposed   for the simple case of uniform and frequency-independent absorptions across all surfaces.  \acp{GA} were also used in \cite{Christensen2014} for the more generalizable task of estimating frequency-dependent absorptions by different surfaces in a room. To do so, measured values of acoustic parameters were matched to simulated ones by adjusting the absorption levels. \acp{GA} are an attractive choice for this problem as the search space for solutions is large and direct minimization of a loss function using a gradient based optimizer is likely to lead to a number of issues. However, the performance of  any method that relies only on audio information to identify materials is limited by the inherent ambiguities in the problem. These ambiguities are due to the fact that many materials can share similar sound-absorption characteristics.    One way to address this   is to  leverage information from other modalities. This was done in \cite{Schissler2018} by using camera images to first detect materials in the room. A \ac{CNN} detector was first trained and used to  initialize a model for the estimation of frequency-dependent absorptions. The estimates were later optimized to match measured \acp{AIR}.  

Computer-vision methods for solving the problem have adopted  state-of-the-art machine learning.  This paper aims to bring state-of-the-art machine learning in material-detection from audio only and to address the ambiguities in the process. A method is proposed  that estimates the probability of presence of   materials in a room. The  method accounts for ambiguities by grouping materials in categories based on the level of sound energy they absorb at different frequencies. Furthermore, instead of attempting to calculate numerical values for the coefficients of each of the unknown number of surfaces in the room, the task is treated as a detection task over the finite number categories. The  detector model is a \ac{CRNN}, which is trained as a multi-task classifier. The training data is created using \textit{a priori} knowledge about the sound absorption properties of various materials. This knowledge is used to create simulated acoustic environments composed of surfaces with specific materials. \acp{AIR} generated from these simulations are presented to the network during training. The network uses information from the reflections encoded in the \acp{AIR} to learn how to detect the materials present.  

This rest of this paper is organized as follows: Section \ref{section_detection_approach} introduces the notation used to model the interaction of sound with material surfaces and presents the proposed method.  The  network's training method, the way that training data is generated  and experiments are given in  Section \ref{section_c5_material_experimetns}.  A discussion and conclusion are given in Section \ref{section_c5_disc_conc}.

\section{Method}
\label{section_detection_approach}

\subsection{Signal Model}
\label{section_notation_and_data}

Considering the case of a perfectly smooth surface, incident sound of frequency $f$  is reflected specularly, resulting in reduced energy and different phase \cite{Kuttruff2009}. The complex factor that represents this process is
\begin{equation}
R(f)=\vert R(f) \vert \exp(i\chi) \text{,}
\end{equation}
with $\chi$ representing the phase difference between the reflected and  incident sound.
The energy of the incident sound absorbed at the surface is described by the absorption coefficient
\begin{equation}
\alpha(f)=1-\vert R(f) \vert ^2 \text{.}
\label{eq_c2_sound_abosrption_coefficient}
\end{equation}
This coefficient is dependent on the frequency of the incident sound.

Assuming $\Theta_{\text{tot}}$ material categories, a category $\theta$ is described by the mean of the absorption coefficients of the materials it describes. 
The coefficients are typically given in 8 1-octave bands for frequencies between 125~Hz and 8~kHz \cite{Ermann2015}. This gives 8 energy absorption coefficients  for each material and material category.  Packing these 8 values together forms column vector $\mathbf{a}_\theta$. Values for   $\Theta_{\text{tot}}$  categories form the matrix of absorption coefficients  $\mathbf{A}_{\text{tot}} = \left[ \mathbf{a}_1, \mathbf{a}_2, \dots, \mathbf{a}_{\Theta{\text{tot}}}   \right]^T$, with $0 < a  < 1 ~\forall~ a \in \mathbf{a}~\forall~ \mathbf{a} \in \mathbf{A}_{\text{tot}}$. 

The  detector model proposed in this paper estimates the probability of presence of materials belonging  to each category.  The matrix of frequency-dependent absorption coefficients of the materials present in an environment is $\mathbf{A}$. Using the presence probabilities, the method described in this paper constructs an estimate $\hat{\mathbf{A}}$  of this matrix   by choosing the appropriate rows of the matrix of known absorptions $\mathbf{A}_{\text{tot}}$. The selected rows will correspond to the categories that are detected as present. Therefore, the aim of the  detector is to perform the function $d$, described as
\begin{equation}
\hat{\mathbf{A}} = d(\mathbf{h}, \mathbf{A}_{\text{tot}}).
\end{equation}

Treating this estimation task as a detection task offers a twofold advantage. It simplifies the problem, as the filter coefficients can be drawn from pre-designed material-filterbanks. It also allows for the use of state-of-the-art detector \acp{DNN} from the literature.

\subsection{Material-category detector \acs{CRNN}}

A \ac{CRNN} model is used as the detector mechanism for the presence of materials in the room. \acp{CRNN} have shown great success in the field of \ac{SED}, as illustrated in the recent \acs{DCASE}~2019 challenge \cite{Kapka2019,MazzonYasuda2019}. In \cite{Papayiannis2018Thesis}, the architecture was also successful in processing \acp{AIR} to categorise individual rooms.  Their success   in \ac{SED} and in classifying reverberant environments motivates their consideration for the detection of materials in this paper.  

As in \ac{SED},  a \ac{CRNN} is trained to estimate the probability of occurrence of an \textit{event} of a certain category. Here, an \textit{event} is considered to be an acoustic reflection. The way that sound energy  is absorbed upon reflection defines the type of event. The \ac{CRNN} is trained as a multi-task classifier, similarly to the task of polyphonic \ac{SED} \cite{Vesperini2019}. This means that after processing a single \ac{AIR} $\mathbf{h}$, the \ac{DNN} returns the posterior probability that  absorption of each of the $\Theta_{\text{tot}}$ categories occurred. Expressing this in the notation introduced in the previous Section, the network   therefore  estimates the  probabilities
\begin{equation}
p( \mathbf{a}_\theta \in \mathbf{A} \vert \mathbf{h} ) ~\forall~ \theta \in \left\{ 1, \dots,  \Theta_{\text{tot}} \right\} \text{.}
\end{equation}
To detect whether a material category $\theta$ is present in the environment, a threshold $\zeta_\theta=0.5$ is applied to the posterior.

\begin{figure}[t]
	\centering
	\includegraphics[scale=0.6]{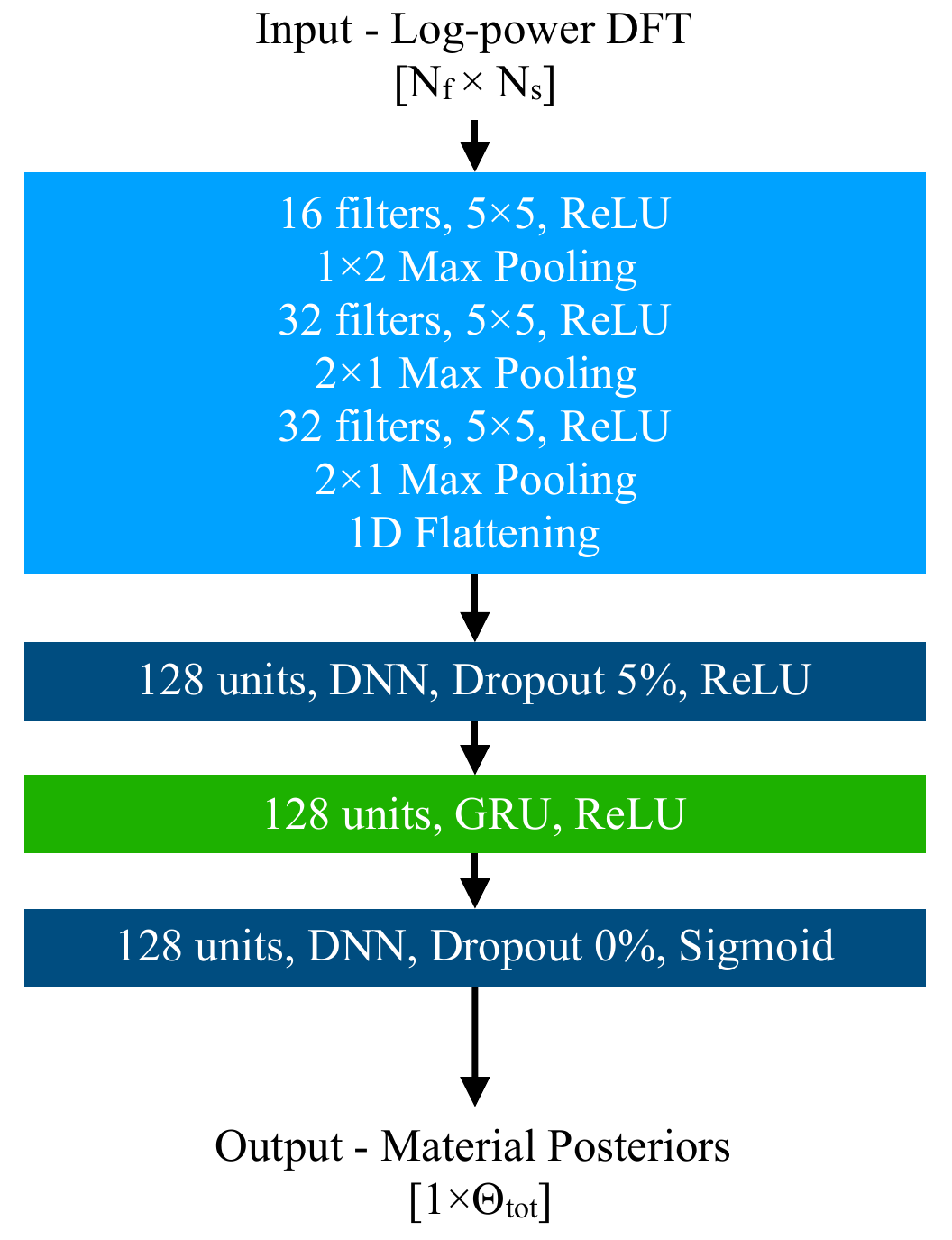} 
	\label{figure_cnn_gru}
	\caption{Proposed \acs{CRNN} for the detection of material categories present in a reverberant acoustic environment, based on their frequency-dependent sound absorption characteristics.}
	\label{figure_material_detection_models}
\end{figure} 

The diagram in Figure \ref{figure_material_detection_models} shows the  network used as the detector. Its inputs are formed using the \ac{FIR} taps of \acp{AIR}. The \ac{FIR} taps  are segmented into frames of duration 3~ms and a 1.5~ms overlap. This provides fine temporal resolution in order to analyze recordings at the reflection level and still a large enough number of samples to maintain significant spectral resolution. The log-power in the discrete frequency domain is presented for each frame at the input, similar to \cite{Papayiannis2018}.

\subsection{Absorption-coefficient data and material clustering}

To train the network, a dataset is needed that is labeled with ground truth information about the materials present in the room. Abortion coefficient tables are available in the literature as acousticians use them as a reference for auralisation experiments and in the design of auditoria. The software package Odeon\footnote{Software homepage: \url{https://odeon.dk/}} is a modeling software that combines such information with acoustic models to create auralisations.  Given the popularity of the software and the fact that a number of manufacturers release their data in a compatible format with it, the data that is available on the software's page\footnote{\label{footnote_chapter5_odeon}The list is freely and publicly available in an electronic format at the time of writing this paper here: \url{https://odeon.dk/sites/all/themes/odeon/images/Materials/Material.Li8}} is used in this work. The  data is used to extract absorption coefficients for 143 materials. The coefficients are given for the 8 1-octave bands in the range 125~Hz to 8~kHz.

As mentioned previously, the proposed method will detect materials in categories. Each category will  represent a subset of the 143 materials.    k-means \cite{Goodfellow2016} is used for creating categories of  materials as clusters.  The Davies-Bouldin criterion \cite{Davies1979} and the \ac{VRC} \cite{Calinski1974} are  used to deduce the number of categories. The absorption coefficients of 143 materials in the 8 1-octave bands are clustered by  the k-means algorithm for a range of  number of clusters between 2--80. The different formulations of the two criteria lead to their optimal values being at opposite extremes. However, the choice of 10 clusters gives a trade-off between the two.  Therefore, the 143 materials are grouped into  $\Theta_{\text{tot}}=10$ categories.

\section{Experiments}
\label{section_c5_material_experimetns}

\begin{table}[t]
\centering
\begin{tabular}{|r|c|ccc|}
\hline
             & Total & Train & Val. & Test  \\ \hline
Percentage   & 100\% & 85\%  & 7.5\%      & 7.5\% \\ \hline
\acsp{AIR} & 70,500 & 59,925 & 5,287       & 5,288  \\ \hline
\end{tabular}
\caption[Partitioning of \acp{AIR} into sets for training and evaluation of \acp{DNN}.]{Partitioning of \acp{AIR} into sets for the training and evaluation of the \ac{CRNN} detector. \acp{AIR} are simulated in shoe-box rooms, using known material frequency dependent absorptions.}
\label{table_c5_air_data_partition}
\end{table}

\begin{table*}[t]
\centering
\begin{tabular}{@{}lrcccccccccc@{}}

\multicolumn{2}{l}{Material Category $\theta$}         & 0     & 1    & 2    & 3    & 4    & 5    & 6    & 7    & 8    & 9    \\ 	\bottomrule
\multicolumn{2}{l}{Materials (out of 143)}                                                                  & 20    & 13   & 12   & 15   & 10   & 3   & 44  & 11   & 19   & 16   \\ 
\multicolumn{2}{l}{Positive Test Samples}                                                                   &  3,080 & 1,738 & 1,573 & 1,564 & 1,910 & 507 & 4,696 & 2,011& 2,495 & 2,497  \\ \bottomrule
\multicolumn{1}{l}{\multirow{3}{*}{\begin{tabular}[c]{@{}l@{}}Baseline SVMs\\IIR  (200,200)  Coef.\end{tabular}}} & Precision                                                          & 0.77   &  0.42 & 0.38   & 0.38   & 0.64   & 0.24   & 0.93   & 0.63  & 0.65   & 0.63      \\ 
\multicolumn{1}{l}{} & Recall                                                           & 0.54    & 0.52   & 0.67   & 0.53   & 0.72   & 0.65   &0.67  & 0.69   & 0.69   & 0.74   \\ 
\multicolumn{1}{l}{} & F1 Score                                                                 & 0.64   &  0.46 & 0.48     & 0.44   & 0.68   & 0.35   & 0.78  & 0.66   & 0.67 & 0.68   \\ \midrule
\multicolumn{1}{l}{\multirow{3}{*}{\begin{tabular}[c]{@{}l@{}}Proposed CRNN\\FIR Taps\end{tabular}}} & Precision                                                          & 0.98    & 0.95   & 0.97   & 0.97   & 0.98   & 0.98   & 1.00  & 0.99   & 0.99   & 0.97   \\ 
\multicolumn{1}{l}{} & Recall                                                           & 0.99    & 0.98   & 0.96   & 0.98   & 0.99   & 0.98   & 1.00  & 0.96   & 0.98   & 0.98   \\ 
\multicolumn{1}{l}{} & F1 Score                                                                 & \textbf{0.98}    & \textbf{0.97}   & \textbf{0.96}   & \textbf{0.97}   & \textbf{0.98}   & \textbf{0.98}   & \textbf{1.00}  & \textbf{0.97}   & \textbf{0.98}   & \textbf{0.98}   \\ \bottomrule
\end{tabular}
\caption{\acs{CRNN} material-category detector  test performance on 5,288 test \acp{AIR} compared to the \ac{SVM}-\ac{IIR} baseline. Positive samples for each material category indicate the number of test \acp{AIR} that contained at least one surface that falls into the specific category. }
\label{table_c5_cnn_performance}
\end{table*}

Simulated \acp{AIR} are used as the training examples for the detector network of Figure \ref{figure_material_detection_models}. The training \acp{AIR} are generated by simulating 141 three-dimensional \textit{shoe-box}  rooms with walls of known frequency-dependent absorptions, using \cite{Wabnitz2010a}.  The size of the simulated rooms is random-uniformly chosen  between [~ 2.5,~2.5,~2.5~] and [~7.0,~7.0,~2.6~]~m. For each one of the 6 walls of the room, the frequency-dependent absorptions are chosen from one of the 143 materials in the list described in Section \ref{section_notation_and_data}.  Each room is populated with 10 sources and 5 receivers at random locations.    Collecting the \ac{AIR} between each source and receiver pair results in 70,500 \acp{AIR}. Each one serves as an in individual training example. The sampling rate used is 16~kHz. The 70,500 generated \acp{AIR}  are split into 3 sets, the training, test and validation set.  Each room contributed \acp{AIR} to only one set. The data partitioning is shown in Table \ref{table_c5_air_data_partition}. Stratified  partitioning is  used
which preserves the positive sample ratios. 

The detector model is trained using the Adam \cite{Kingma2014} optimizer with a cross-entropy loss.  The batch size is set to 128 \acp{AIR} of duration 200~ms, which are split into frames of 3~ms with 1.5~ms overlap. Overfitting is prevented by early stopping which stops the training of the model 10 epochs after the training loss stopped improving or 15 epochs after the validation loss stopped improving.  The final model is selected at the epoch with the minimum validation loss.  Since the ratios of positive samples are not even across each of the 10 material categories, the contribution to the cross-entropy loss of each \ac{AIR}  is weighted  as proposed in \cite{Pedregosa2011}.

As a baseline, a set of \acp{SVM} are trained to perform the same task. This compares the use of the proposed end-to-end \ac{CRNN} for the detection with a  feature-based classifier. For this baseline, one \ac{SVM} is trained per material category to make the binary decision of material presence or not. Training the 10 \acp{SVM} using the \ac{AIR} \ac{FIR} taps is impractical due to their high dimensionality, To address this, an alternative \ac{AIR} representation is considered.  \ac{IIR} models offer a  parsimonious representation of \acp{AIR} \cite{Karjalainen2002a} that can capture information regarding resonances and frequency-regions of sound absorption in a room. Given their relevance to the task and their low-dimensionality, the coefficients of \ac{IIR} models of \acp{AIR} are then used as the feature-vector inputs to the \acp{SVM}. The number of coefficients in the numerator and denominator are 200 each.  This choice gave a trade-off between accuracy and training times. Further increasing the number of coefficients did not significantly improve the results but significantly increased training times. The \ac{FIR} taps of \acp{AIR} were used to derive the \ac{IIR} coefficients using Prony's method \cite{Parks1987}. When training the \acp{SVM}, the samples were weighted to counter imbalances and make the comparison fair.

The detection performance of the model and the baseline is measured using the harmonic mean of precision and recall, the $F_1$ score.

The result of evaluating the \acs{CRNN} based on the above measures is given in Table  \ref{table_c5_cnn_performance}.  The F1 score for the network's predictions is 96\% for the worst performing category and 100\% for the best.  In terms of individual material types, unsurprisingly the best results are obtained for $\theta=6$, which is the biggest cluster, containing $63.9\%$ of the training examples. The baseline \ac{SVM} detector models show worst performance across all categories despite taking proportional training times to the \ac{CRNN} model. Despite the class weighting being applied to both the baseline and the \ac{CRNN}, the baseline detectors for minority categories show significantly lower $F_1$ scores. The low precision of the \acp{SVM} for those categories indicate a high-number of false positives, which make this baseline a far less attractive option when compared to the proposed \ac{CRNN}.

\section{Discussion and conclusion}
\label{section_c5_disc_conc}

This paper proposed a novel method for estimating the probability of  presence of material categories in a room, based on their  sound absorption characteristics. The method takes as input  the \ac{FIR} filter taps of one \ac{AIR}. It does not assume visual access to the room, which is the case for previously proposed methods \cite{Schissler2018}. The detection is performed by a \ac{CRNN} that is trained as a multi-task classifier. 
In the experiments presented in this paper, the  material detector network was tested on more than 5,000 simulated  \acp{AIR} and 143 materials. The F1 score for the network's predictions in the tests was  96\% for the worst performing category and 100\% for the best. The model was compared to a set of \ac{SVM} detectors, which relied on the \ac{IIR} representation of \acp{AIR}. Comparing this with the proposed \ac{CRNN} that was inspired by \ac{SED}, shows that the $F_1$ for the proposed model is on average 40.3\% higher. Therefore, allowing a \ac{DNN} to process short frames of the \acp{AIR} and treat reflections as indicators for acoustic events outperforms  methods that rely on aggregate representations of the spectrum of the \ac{AIR}.

The detected material categories allow for the estimation of the frequency-dependent absorption coefficients of the surfaces in an acoustic environment.   This is done by simply processing a single \ac{AIR} measured in the room. This is in contrast to other methods that require the extraction of samples of  materials from the environment, such as  \acs{ISO}-354   and \acs{ISO}-10534. This  estimation allows for the reconstruction of the acoustics of a given room \cite{Schissler2018}. Furthermore, using the estimates, the parametric modeling of \acp{AIR} becomes more accurate  and improves the estimation of the \acp{TOA} of acoustic reflections \cite{Papayiannis2018Thesis}. This leads to a cascade of other applications, such as room geometry estimation \cite{Moore2013}.

\bibliographystyle{ieeetr}
\bibliography{ms}

\end{document}